\documentclass[conference]{IEEEtran}
\IEEEoverridecommandlockouts

\usepackage{cite}
\usepackage{amsmath,amssymb,amsfonts}
\usepackage{algorithmic}
\usepackage{graphicx}
\usepackage{textcomp}
\usepackage{xcolor}
\def\BibTeX{{\rm B\kern-.05em{\sc i\kern-.025em b}\kern-.08em
    T\kern-.1667em\lower.7ex\hbox{E}\kern-.125emX}}
\begin{document}

\title{Retrieval Augmented Correction of Named Entity Speech Recognition Errors}

\author{
\IEEEauthorblockN{Ernest Pusateri\IEEEauthorrefmark{1}\IEEEauthorrefmark{3}, Anmol Walia\IEEEauthorrefmark{1}\IEEEauthorrefmark{3}, Anirudh Kashi\IEEEauthorrefmark{1}\IEEEauthorrefmark{3}, Bortik Bandyopadhyay\IEEEauthorrefmark{1}\IEEEauthorrefmark{3}, Nadia Hyder\IEEEauthorrefmark{1}\IEEEauthorrefmark{3}, \\
Sayantan Mahinder\IEEEauthorrefmark{3},
Raviteja Anantha\IEEEauthorrefmark{3},
Daben Liu\IEEEauthorrefmark{2}\IEEEauthorrefmark{4} and Sashank Gondala\IEEEauthorrefmark{2}\IEEEauthorrefmark{5}}
\IEEEauthorblockA{\IEEEauthorrefmark{3}Apple \\
Email: \{epusateri,awalia,a\_kashi,bbandyopadhyay,n\_hyder\}@apple.com
}
\IEEEauthorblockA{\IEEEauthorrefmark{4}Capital One}
\IEEEauthorblockA{\IEEEauthorrefmark{5}Further AI}
\thanks{\IEEEauthorrefmark{1} Equal contribution.}
\thanks{\IEEEauthorrefmark{2} Work done at Apple.}
}

\maketitle

\begin{abstract}

In recent years, end-to-end automatic speech recognition (ASR) systems have proven themselves remarkably accurate and performant, but these systems still have a significant error rate for entity names which appear infrequently in their training data.  In parallel to the rise of end-to-end ASR systems, large language models (LLMs) have proven to be a versatile tool for various natural language processing (NLP) tasks.  In NLP tasks where a database of relevant knowledge is available, retrieval augmented generation (RAG) has achieved impressive results when used with LLMs.  In this work, we propose a RAG-like technique for correcting speech recognition entity name errors.  Our approach uses a vector database to index a set of relevant entities. At runtime, database queries are generated from possibly errorful textual ASR hypotheses, and the entities retrieved using these queries are fed, along with the ASR hypotheses, to an LLM which has been adapted to correct ASR errors.  Overall, our best system achieves 33\%-39\% relative word error rate reductions on synthetic test sets focused on voice assistant queries of rare music entities without regressing on the STOP test set, a publicly available voice assistant test set covering many domains. 

\end{abstract}
%
%
\section{Introduction}
\label{sec:intro}

In recent years, end-to-end automatic speech recognition (ASR) systems have risen in prominence, as they have proven themselves remarkably accurate and performant \cite{prabhavalkar2023end}. However, these systems still have a significant error rate for entity names that appear infrequently in their training data \cite{van2023modeling}. In parallel to the rise of end-to-end ASR systems, large language models (LLMs) have proven to be versatile tools for various natural language processing (NLP) tasks, including ASR error correction \cite{ma2023can}. In NLP tasks where a database of relevant knowledge is available, retrieval augmented generation (RAG) approaches have been especially effective when used with LLMs, e.g. \cite{izacard2023atlas}. In RAG-based approaches, information from an external knowledge base is retrieved and provided to the LLM as part of the input context.  RAG-based approaches allow LLMs, especially those with fewer parameters, to perform better on tasks requiring a large amount of domain specific knowledge \cite{kandpal2023largelanguagemodelsstruggle}.

In this work, we propose a RAG-like technique for correcting speech recognition entity name errors. Our approach uses a vector database to index a set of relevant entities. At runtime, database queries are generated from possibly errorful ASR hypotheses, and the entities retrieved using these queries are fed, along with the ASR hypotheses, to an LLM which has been adapted to correct ASR errors. Importantly for correcting ASR errors, the encodings used as queries and database keys capture information about the likely acoustics of the corresponding text strings.

The main contributions of this work are a RAG-based technique for correcting entity name errors in textual ASR hypotheses using a large entity database, as well as an evaluation of strategies for retrieving relevant entities, generating queries and adapting an LLM within the proposed technique.

\section{Approach}
\label{sec:approach}

\begin{figure*}[htp]
    \centering
    \includegraphics[width=18cm]{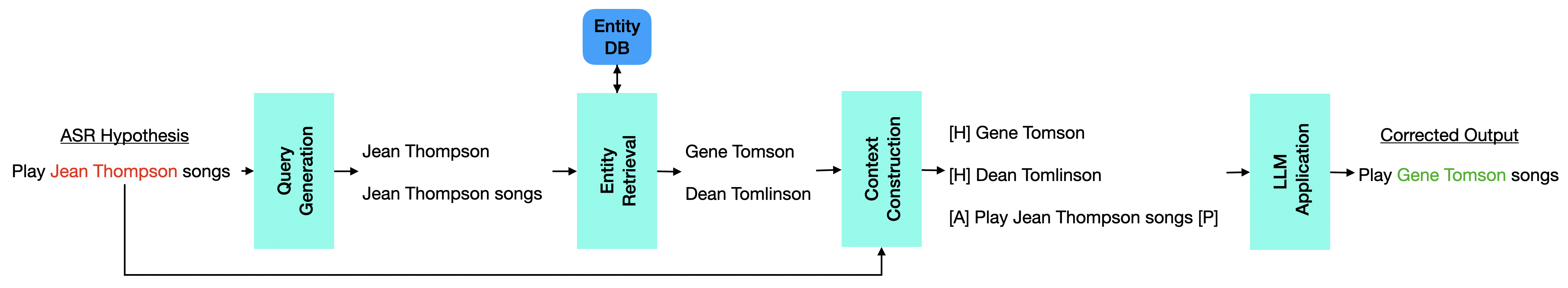}
    \caption{Our RAG-based approach. with an example. ASR errors are in red and corrected errors are in green. In the context, [H] precedes each hint, [A] precedes the ASR transcript and [P] is the cue for the LLM to start prediction.}
    \label{figure:approach}
\end{figure*}

Our RAG-based approach uses a vector database to index the entities and retrieve them during runtime. Rather than querying the database using the full transcript, we use a query generation process to derive one or more queries. Hence, our approach consists of four steps, as illustrated in Figure \ref{figure:approach}: Query Generation, Entity Retrieval, Context Construction and LLM Application.  We describe these steps in the following sub-sections.

\subsection{Query Generation}
\label{ssec:query_gen}

Our goal in this step is to generate queries from the ASR hypothesis that are likely to result in relevant entities being retrieved.  We explore three approaches, each with its own tradeoffs.

\textit{All N-grams}: In this approach, all subword sequences in the hypothesis up to an empirically determined length, $N_{max}$ are used as queries. This approach has the advantage that it requires no separate model or set of rules. However, it will over-generate, which may make the task of using the retrieved entities for correction more difficult.

\textit{Template Matching}: Here we use a hand-curated domain-inspired set of regular expressions to extract regions that likely correspond to named entities.  For instance, \texttt{/play (.*)/} might be used for the music domain.  This approach has an advantage over the \textit{All N-grams} in that it will produce fewer queries.  However, it requires the manual creation of a set of regular expressions, and these regular expressions will inevitably miss some edge cases.

\textit{NE Tagging}: In this approach we train a separate named entity region tagger to mark regions of the ASR output likely to correspond to named entities. As with \textit{Template Matching}, this approach will produce fewer queries than \textit{All N-grams}.  Unlike \textit{Template Matching}, this approach does not require creation of a set of regular expressions. It is also more likely to generalize to edge cases. This approach requires the training and application of a separate tagging model, which introduces complexity and increases inference time.

\subsection{Entity Retrieval}
\label{ssec:entity_ret}

For retrieval, we construct a vector database from a list of named entities using one of three methods for query and key generation: \textit{Okapi BM25} \cite{robertson1995okapi} and \textit{T5 Semantic Embeddings} \cite{ni2021sentence}, which capture semantic similarity, and \textit{Acoustic Neighbor Embeddings} \cite{jeon2020acoustic}, a type of acoustic word embedding. \textit{Acoustic Neighbor Embeddings} are generated either from the orthography or pronunciation of a word sequence and are optimized to capture the likely acoustics of the word sequence. We evaluate generating embeddings from both orthography and pronunciation.  

\subsection{Context Construction and LLM Application}
\label{ssec:context_construction}
To minimize compute and memory requirements without compromising accuracy, we want to keep the context size small while ensuring that relevant entities are included.  To accomplish this, we use a two-step context generation process. We first filter the retrieved entities and then construct a \textit{hint} for each of the remaining entities. To filter, we apply two thresholds, 1. $D_{max}$, as a Euclidean distance threshold for keeping the most relevant candidates and a count threshold, 2. $R_{max}$ to limit the overall candidate counts. 

We experiment with two different hint string formats: one containing only the retrieved entity, and the other containing the retrieved entity plus the search query used to retrieve the entities. The LLM input is constructed by appending the ASR hypothesis to the hint string. We use tags to delineate the the hints and ASR hypothesis, as shown in Figure \ref{figure:approach}. Once the context is constructed, it is provided as input to an adapted LLM.

\section{Related Work}
\label{sec:related_work}
There has been a great deal of previous work on the problem of ASR error correction in general, e.g. \cite{guo2019spelling,gu2024denoising}, some of which was focused on correcting named entity errors in particular \cite{raghuvanshi2019entity,Wang2020ASREC,kang2024transformer}. These entity-focused works include phonetic input representations to help the models leverage acoustic confusability, but none of these approaches include a retrieval step, which means they are limited by the model’s ability to memorize relevant entities.

In work more similar to ours, \cite{cai2023kg} presents a query rewriting approach that can leverage acoustic confusability and includes a database retrieval step from a large entity database.  That work differs from ours in that the entire text input is encoded for entity retrieval from the database using a task-specific encoder, and correction is done using a task-specific model architecture. In contrast, our system is simpler, in that we use task-independent encoders and our model is a small LoRA adapter with a pre-trained LLM. The approach described in \cite{zhang2023knowledge} is also similar to ours, but rather than an employing a adapted LLM for error correction, ASR is rerun with the generated context, making it more computationally costly. In \cite{ragdst}, a RAG-like approach is applied to a dialog state tracking task. The database used in that work had only 2500 entities whereas we use a database of 2.6 million entitites. While our approach starts from a textual ASR hypothesis, both \cite{zhang2023knowledge} and \cite{ragdst} requre the input audio, which means they cannot be applied in scenarios where only the textual ASR hypothesis is available.

\section{Experimental Design}
\label{sec:experiments}

\subsection{Data}
\label{ssec:data}
To evaluate our approach, we measure it's effect in a scenario where we wish to improve accuracy on music entities within a voice assistant (VA) system without degrading accuracy in other domains. To construct this scenario, we use two data sources.  To represent the general voice assistant task, we use the STOP dataset \cite{tomasello2023stop}, which includes human and synthetic audio recordings along with transcripts in eight VA domains.  This dataset also includes semantic analyses, which we use to infer music entity spans. Music queries in this set account for only 9.2\% of utterances and cover mostly common music entities.

To cover a larger number of rare music entities, both for training and evaluation, we synthetically generated music queries using templates and entity lists as detailed in \cite{van2022space}. The synthetic queries were stratified according to rank percentiles, following the methodology outlined in \cite{van2022space}, and we sampled training, validation and evaluation sets from each stratified partition (head, torso, tail). The corresponding audio was synthesized using a neural Text-to-Speech (TTS) system as described in \cite{achanta2021device}. To ensure the integrity of the evaluation process, we constructed training and evaluation sets with non-overlapping entities and employed different TTS voices for audio generation.

The sizes of the training and evaluation datasets from each data source are presented in Table \ref{table:ds}. The entity catalog from \cite{van2022space}, consisting 2.6 million entities, was employed without modification in experiments involving the STOP dataset. All datasets are decoded using an in-house Connectionist Temporal Classification (CTC)-based end-to-end ASR system, following \cite{zhang2022wenet}, \cite{lei2024personalization}, \cite{lei2023acoustic}. This system incorporates two word-level external language models (LMs): a neural LM trained on anonymized usage query logs and a 4-gram LM trained on synthetic queries, consistent with the methodology described in \cite{van2022space}.

\begin{table}[h]
\caption{Dataset sizes}
\centering
\begin{tabular}{|l|c|c|c|}
\hline
& \multicolumn{3}{c|}{\# of utterances} \\
\cline{2-4}
 & train & validation & eval \\
\hline
head & 98848 & 9913 & 9894 \\
torso & 96722 & 9667 & 9678 \\
tail & 95485 & 9516 & 9563 \\
stop & 120929 & 33387 & 75640 \\
\hline
\end{tabular}
\label{table:ds}
\end{table}
\subsection{Models}
\label{ssec:models}
As mentioned in Section \ref{ssec:entity_ret}, we use the acoustic word embeddings proposed in \cite{jeon2020acoustic}, \textit{Acoustic Neighbor Embeddings}, as one approach to generate queries and keys used in the Entity Retrieval step. This approach involves training acoustic, phonetic and orthographic encoders to minimize the Euclidean distance between acoustic, phonetic and orthographic embeddings of the same phrase. Importantly for our purposes, this also has the effect of minimizing the Euclidean distance between encodings of phrases with similar acoustic realizations. In this work, we use 40-dimensional embeddings.

For the generation task we used OpenLLaMA \cite{openlm2023openllama} 7B model from the LLaMA \cite{touvron2023llama} family. For NE tagging in query generation as mentioned in section \ref{ssec:query_gen}, we separately adapt the OpenLLaMA \cite{openlm2023openllama} model using LoRA adapters \cite{hu2021lora} with rank 16, employing the same hyperparameters as detailed in \ref{ssec:models}. To be consistent with the synthetic data, all music entity types in the STOP data are consolidated into a single type.

\subsection{Hyperparameters}
\label{ssec:models} 
We set the distance threshold $D_{max}$ to 1.0 in all experiments. $N_{max}$, the maximum query length when using the all-ngrams query generation, is set to $5$. We experiment with $R_{max}$ values of $1$ and $5$.

All experiments use the AdamW optimizer \cite{loshchilov2017fixing} for LLM adaptation, with $\alpha = 0.001$, $\beta_1 = 0.9$, $\beta_2 = 0.999$, $\epsilon = 1e-8$ and $\lambda = 0.001$.  We use a cosine learning rate scheduler with a linear warmup of 100 steps. In LoRA experiments, we use 2000 training steps, whereas in full model fine tuning, we use 4000 training steps. In all experiments, we use an effective batch size of 128. For LLM inference, greedy decoding with 1000 maximum tokens is used in all experiments to minimize computational and memory overhead.

\section{Results}
\label{sec:results}

We assess the overall efficacy of our approach using the standard ASR WER metric, quantifying the discrepancy between the fine-tuned LLM output and reference transcriptions. We start by evaluating methods for key and query vector generation for the Entity Retrieval step. For this purpose, we measure recall of the top 1, top 5, and top 10 closest matches for each method on the synthetic test sets. As text queries for retrieval, we use the (possibly errorful) span of words in the ASR output corresponding to an entity, extracted using the Template Matching approach. The results of this evaluation can be seen in Table \ref{table:rt}. We first observe that using orthography or phoneme \textit{Acoustic Neighbor Embeddings} achieves much higher recall than either using \textit{T5 Semantic Embeddings} or \textit{Okapi BM25}. This is consistent with the intuition that in this task we are interested in acoustic similarity rather than semantic similarity. 

\begin{table}[h]
\caption{Comparison of different retrieval methods.}
\centering
\begin{tabular}{|l|c|c|c|}
\hline
& \multicolumn{3}{c|}{Recall (top 1/top 5/top 10) (\%)} \\
\cline{2-4}
\multicolumn{1}{|c|}{} & head & torso & tail \\
\hline
Okapi BM25 & 53.5/71.2/74.6 & 56.6/71.5/74.1 & 59.5/72.6/74.9 \\
T5 & 80.7/86.9/88.4 & 78.6/85.6/87.3 & 78.7/85.8/87.6 \\
AN-Ortho & 84.8/91.6/92.8 & 83.6/90.3/91.6 & 83.9/90.1/91.4 \\
AN-Phoneme & \textbf{85.6/92.1/93.3} & \textbf{84.1/90.7/92.1} & \textbf{84.4/90.7/92.0} \\
\hline
\end{tabular}
\label{table:rt}
\end{table}

Focusing on \textit{Acoustic Neighbor Embeddings}, we observe a large gain in recall when we compare using the top 5 closest matches to the top 1.  However, increasing the number of matches further to 10 results in a less than 2\% increase in recall. Comparing using \textit{Acoustic Neighbor Embeddings} derived from orthography to those derived from phonemes, we see that using phonemes achieves less than 1\% higher recall while having the additional requirement of a G2P model or lexicon. Based on these results, we chose orthography-based \textit{Acoustic Neighbor Embeddings} for our experiments and evaluated only using the best matches in the top-1 and top-5.

Next we evaluate different strategies for adapting the LLM to our task: LoRA adapters \cite{hu2021lora} with 4 different ranks and full fine-tuning.   In these experiments, we use the \textit{All N-gram} query generation method, and we do not include the query in the hints. Results are shown in Table \ref{table:ts}. We see that LoRA at all ranks and full fine-tuning achieve similar WERs. We also experimented with in-context learning (ICL), but this strategy resulted in large WER regressions. As rank 4 LoRA requires the fewest task-specific parameters of the best performing strategies, we use it for the remainder of our experiments. 

\begin{table}[h]
\caption{Comparison of different training strategies.}
\centering
\begin{tabular}{|l|c|c|c|c|}
\hline
& \multicolumn{4}{c|}{WER (\%)} \\
\cline{2-5}
\multicolumn{1}{|c|}{} & head & torso & tail & STOP \\
\hline
LoRA-4 & 4.68 & 5.02 & 4.97 & 4.27 \\
LoRA-8 & \textbf{4.66} & \textbf{5.02} & \textbf{4.93} & \textbf{4.23} \\
LoRA-16 & 4.67 & 5.05 & 4.93 & 4.24 \\
LoRA-32 & 4.80 & 5.11 & 5.1 & 4.65 \\
Full fine-tuning & 4.74 & 5.11 & 5.03 & 4.48 \\
\hline
\end{tabular}
\label{table:ts}
\end{table}

Next, we evaluate different methods for the Query Generation step.  The results of these experiments are shown in Table \ref{table:qg}.  We’ll first consider the result for no entity hints, which represents the scenario where the LLM is provided with no external information to correct the ASR hypothesis. Compared to the baseline system with no error correction, we see a relative WER improvement of 1-2\% on synthetic sets and 5.7\% on the STOP set. The greater improvements on the STOP set can be attributed to its inclusion of queries from diverse domains, which are more amenable to grammatical or formatting corrections, unlike the synthetic sets created with short templates. For example, the LLM effectively corrects a common ASR error where an ordinal date is misrecognized as its cardinal form (e.g., 'Thirteenth' versus 'Thirteen'). 

Moving on to experiments with entity hints, we evaluate each query generation method with 1 match per query and 5 matches per query. As shown in Table \ref{table:qg}, all query generation approaches demonstrate significant WER improvements in synthetic sets, achieving a relative reduction of WER of up to 39\% in the tail set. The improvements observed on the STOP set are less pronounced and, upon deeper inspection, primarily due to correction of non-music-related words.  This is expected, given that the number of music queries in the STOP dataset is only 9.2\%, and these queries are heavily weighted toward more common music entities on which the speech recognizer makes few errors.  However, the small improvements in WER on the STOP dataset do demonstrate that our approach does not degrade performance on average on queries outside the music domain or on music queries that are already correctly recognized. Among the query generation variants, \textit{Template Matching} and \textit{NE Tagging} outperform \textit{All N-gram}, albeit marginally. The \textit{All N-gram} method, while slightly less effective, offers a straightforward approach to query generation eliminating the need for separate model training or template curation.

Incorporating multiple matches per query offers minimal improvement for \textit{NE Tagging} and \textit{Template Matching}, while increasing computational overhead during inference. However, it leads to regression in \textit{All N-gram}, as the additional matches seem to confuse the model by introducing more incorrect candidates into the context, compounded by the already increased number of queries in this approach. The lack of significant improvement when using multiple matches suggests that the adapted LLM lacks any intelligence that would help distinguish between acoustically similar entities, making the best strategy to consider only the most acoustically similar entity for any particular query.

\begin{table}[h]
\caption{Comparison of different query generation approaches.}
\centering
\begin{tabular}{|l|l|c|c|c|c|}
\hline
& & \multicolumn{4}{c|}{WER (\%)} \\
\cline{3-6}
$R_{max}$ & Generation Strategy & head & torso & tail & STOP \\
\hline
- & No correction & 6.98 & 7.68 & 7.98 & 4.53 \\
- & No hints & 6.9 & 7.58 & 7.84 & 4.27 \\
1 & All N-grams & \textbf{4.68} & 5.02 & 4.97 & 4.27 \\
1 & Template Matching & 4.7 & \textbf{4.97} & \textbf{4.87} & \textbf{4.18} \\
1 & NE Tagging & 4.75 & 5.0 & 4.92 & 4.24\\
5 & All N-grams & 4.83 & 5.08 & 5.02 & 4.31 \\
5 & Template Matching & 4.69 & 5.0 & 4.89 & 4.2 \\
5 & NE Tagging & 4.7 & 5.0 & 4.9 & 4.19 \\
\hline
\end{tabular}
\label{table:qg}
\end{table}

Finally, we evaluate the effect of including the queries in the hints. These results are shown in Table \ref{table:cc}. Here we use the \textit{All N-gram} generation strategy, which seems likely to benefit most from query inclusion, as, in this strategy, hints are generated for multiple ASR hypothesis spans.  We observe little impact on WER which suggests the model is able to easily infer the spans in the ASR hypothesis for which the retrieved entities are relevant.

\begin{table}[h]
\caption{Effect of including query text in the context.}
\centering
\begin{tabular}{|l|l|c|c|c|c|}
\hline
& & \multicolumn{4}{c|}{WER (\%)} \\
\cline{3-6}
$R_{max}$ & Query Included & head & torso & tail & STOP \\
\hline
1 & & \textbf{4.68} & 5.02 & 4.97 & \textbf{4.27} \\
1 & \checkmark & 4.69 & \textbf{5.01} & \textbf{4.94} & 4.29 \\
5 & & 4.83 & 5.08 & 5.02 & 4.31 \\
5 & \checkmark & 4.81 & 5.08 & 4.98 & 4.33 \\
\hline
\end{tabular}
\label{table:cc}
\end{table}

Overall, the best system achieves 32.6\%, 35.3\%, 39.0\% and 7.7\% relative WER reductions on the head, torso, tail and STOP test sets respectively over doing no error correction.  

\section{Conclusion}
\label{sec:conclusion}

In this work, we presented a RAG-based technique for correcting entity name errors in ASR hypotheses using an adapted LLM. We evaluated different embedding methods to use for entity retrieval and found that Acoustic Neighbor Embeddings significantly outperformed semantic embeddings, achieving entity recall rates more than 4\% higher.  We also showed that using a LoRA adapter with rank 4 works as well as higher rank LoRA adapters or full fine-tuning on this task when using the 7B parameter OpenLLaMA LLM. Our best system achieved 33\%-39\% relative WER reductions on synthetic test sets focused on voice assistant queries of rare music entities without regressing
on the STOP test set which covers many domains.

\bibliographystyle{IEEEtran}
\bibliography{strings,refs}

\end{document}